\magnification=1200
\parindent=30pt
\parskip 10pt plus 1pt
\rm
\baselineskip 12pt
\null

\centerline{\bf NEW IMPROVEMENTS FOR MIE SCATTERING CALCULATIONS}
\vskip .3in
\centerline{V. E. Cachorro}
\centerline{Departamento de F\'{\i}sica Aplicada I}
\centerline{Valladolid University, 47071 Valladolid, SPAIN}
\vskip 0.3in
\centerline{L. L. Salcedo}
\centerline{Departamento de F\'{\i}sica Moderna}
\centerline{Granada University, 18071 Granada, SPAIN}
\null
\vskip 0.5in
\baselineskip 12pt
\centerline{\bf ABSTRACT}
New improvements to compute Mie scattering quantities are presented. They 
are based  on a detailed analysis of the various sources of error in 
Mie computations and on mathematical justifications. The algorithm 
developed on these improvements  proves to be reliable 
and efficient, without size ($x=2\pi R/\lambda$) nor refractive index 
($m=m_R-{\rm i}m_I$) limitations, and the user has a choice to 
fix in advance the desired 
precision in the results. It also includes a new and efficient method to 
initiate the downward recurrences of Bessel functions.
\vfill
\eject

{\bf 1. INTRODUCTION}

The Mie theory of light scattering by a homogeneous sphere
is used for many problems of 
atmospheric optics and also in other fields in Physics. The application 
of Mie theory still needs modern computers for numerical calculations
of the many functions and coefficients involved. The primary difficulty 
is in the precise evaluation of expansion coefficients $a_n$ and $b_n$. 
This is further aggravated as $x$ gets large, and when the calculation of 
size distribution is needed. An optimization of computer time for 
reliable computation is clearly of necessity.

The formulas for Mie scattering are well known${}^{1,2}$. Here we follow 
the notation of  Bohren and Huffman${}^{3}$. 
The scattering and extinction efficiency factors are given by
$$\eqalign{Q_s &= {2\over x^2}\sum^N_{n=1}(2n+1)\big(|a_n|^2 +
|b_n|^2\big)\cr
Q_e &={2\over x^2}\sum^N_{n=1}(2n+1){\rm Re}\,(a_n+b_n)\cr}\eqno(1)$$
where $x=2\pi R/\lambda$ is the size parameter of the problem, $R$ 
being the radius of the sphere, $\lambda$ the wavelength of the 
light and $N$ a large enough number. The Mie scattering  coefficients 
$a_n$ and $b_n$ are functions of 
$x$ and the relative refractive index $m=m_R-{\rm i} m_I$, with $m_R\geq 
1,\,m_I\geq 0$.
$$\eqalign{ a_n &= 
{x\psi_n(x)\psi^\prime_n(y) -y\psi^\prime_n(x)\psi_n(y)\over
x\zeta_n(x)\psi^\prime_n(y) -y\zeta^\prime_n(x)\psi_n(y)}\cr
 b_n &= 
{y\psi_n(x)\psi^\prime_n(y) -x\psi^\prime_n(x)\psi_n(y)\over
 y\zeta_n(x)\psi^\prime_n(y) -x\zeta^\prime_n(x)\psi_n(y)}\cr}\eqno(2)$$
where $y=mx$ and $\psi_n(z), \zeta_n(z)$ are the Riccati-Bessel functions 
related to the spherical Bessel functions $j_n(z)$ and 
$y_n(z)$:
$$\eqalign{\psi_n(z) &= z\,j_n(z) \cr
\zeta_n(z) &= z\,j_n(z) - {\rm i}z\,y_n(z)   \cr}\eqno(3)$$
These functions are known in closed form (Ref. 4, p. 437) but it is 
more convenient to use the recurrence relation
$$\eqalign{X_{n+1}(z) &= F_n(z) X_n(z) - X_{n-1}(z),\cr
  F_n(z) &= (2n+1)/z\ .\cr}\eqno(4)$$
where $X$ is any of the functions in eqn. (3). 

Presently, there are many versions of Mie scattering computer codes 
(Dave${}^{5,6}$, Blattner$^7$, Grehan and Gouesbet$^{8,9}$, 
Wiscombe$^{10,11}$,
Goedecke et al.$^{12}$, Miller$^{13}$) and authors who had been doing Mie 
calculations (Kattawar and Plass$^{14}$, Deirmendjian$^{15}$, Quenzel and 
M\"uller$^{16}$, Bohren and Huffman$^{3}$). These are reflected in 
performing our work.

Some essential points should be addressed by any Mie scattering algorithm:
\item{1)} How to determine the number $N$ for truncating a Mie series.
\item{2)} Whether the Riccati-Bessel functions will be computed by
          upward recursion or by downward recursion.
\item{3)} If downward recursion is used, how to initialize it.
\item{4)} How to structure the algorithm in an efficient way.

Answers to all the above questions constitute the objective of this 
paper. We focus particularly on analyzing the numerical error sources 
and show that our Mie algorithm permits users to prescribe a precision 
$\epsilon$ beforehand, to effect an efficient, reliable Mie coefficients 
calculation. Needless to say, the precisely evaluated Mie coefficients 
$a_n,\,b_n$ are required for calculating the angular scattering 
amplitudes${}^{1,2,3,5,6,10}$.

{\bf 2. CONVERGENCE PROPERTIES OF THE MIE SERIES}

In this section we shall estimate the error introduced in some typical 
quantity such as the efficiency factors, by keeping a finite number $N$ 
of partial waves in the Mie series. We shall also find a criterion for 
choosing the value of $N$. In this section the quantities 
$a_n, b_n$ themselves are assumed to be computed exactly.

In order to investigate the convergence properties of the scattering 
coefficients $a_n, b_n$ we shall make use of very well known  properties 
of the spherical Bessel functions (e.g. ref. 4, p. 438 and ff.). Let us 
recall some properties which are relevant for us:
\item{i)}
$$\displaystyle\lim_{n\to\infty}\psi_n(z) = 0\,, \ \ \ 
\displaystyle\lim_{n\to\infty}\zeta_n(z)=\infty\,. \eqno(5)$$
\item{ii)} For $z=x$ real, $\psi_n(x)$ and $\zeta_n(x)$ have two distinct 
regimes as functions of $n$:
\item{a)} oscillating regime for $n<x$. $\psi_n(x)$ and $\zeta_n(x)$ 
keep changing their sign regularly, and $|\psi_n(x)|$ and $|\zeta_n(x)|$ 
are bounded by slowly changing functions of $n$.
\item{b)} exponential regime for $n>x$. $\psi_n(x)$ becomes 
exponentially decreasing and $|\zeta_n(x)|$ becomes exponentially 
increasing.

In view of these considerations one concludes from eqn. (2), that all 
the partial waves $n<x$ ($x$ being the size parameter from now on) will 
contribute to the Mie series and convergence will appear only after $n$ 
enters in the exponential regime. This is so because 
$\psi_n(x),\psi^\prime_n(x)$ go very quickly to zero in the numerator 
and $\zeta_n(x),\zeta_n^\prime(x)$ go to infinity in the denominator. On 
the other hand $\psi_n(y),\psi^\prime_n(y)$ appear both in numerator and 
denominator and therefore seem to play no role in the convergence.
We can emphasize this fact by writing
$$\eqalign{a_n &= {\psi_n(x)\over\zeta_n(x)}\,[a]_n 
= {\psi_n(x)\over\zeta_n(x)}\, {n(y/x-x/y)+xA_n(y)-yA_n(x)\over
n(y/x - x/y) + xA_n(y)-yB_n(x)} \cr
b_n &= {\psi_n(x)\over\zeta_n(x)}\,[b]_n
= {\psi_n(x)\over\zeta_n(x)}\, {yA_n(y)-xA_n(x)\over
yA_n(y)-xB_n(x)} \cr}\eqno(6)$$
where we have extracted the factor $\psi_n(x)/\zeta_n(x)$ responsible 
for the convergence of $a_n$ and $b_n$ and also we have reexpressed the 
ratios $\psi^\prime_n(z)/\psi_n(z)$ and $\zeta^\prime_n(x)/\zeta_n(x)$ 
in terms of (ref. 4, p. 439)
$$A_n(z)={\psi_{n-1}(z)\over\psi_n(z)}\ ,\ \ \ \ \ 
B_n(x)={\zeta_{n-1}(x)\over\zeta_n(x)}\eqno(7)$$

Let us state more clearly our assumption: we shall assume that the 
quantities $[a]_n,[b]_n$ are bounded by slowly varying functions of $n$ 
in the exponential regime $n>x$. The validity of this assumption will be 
analyzed in a later section.

If $[a]_n$ and $[b]_n$ are well behaved for large $n$, we can 
approximate them by their asymptotic values in order to discuss the 
convergence of $a_n$ and $b_n$. In order to take advantage of this 
approximation we can use the asymptotic expansion of the Bessel 
functions for large orders (ref. 4, p. 365),
$$A_n(z)\sim F_n(z),
\ \ \ \ B_n(x)\sim F^{-1}_n(x)\ \ \ \ {\rm as }\ \ n\to\infty\eqno(8)$$
where the next term in the expansion has a higher power of $1/n$. We 
obtain
$$[a]_n\sim {1-m^2\over 1+m^2}+O\big({1\over n}\big),\ \ \ \ \ 
  [b]_n\sim O\big({1\over n}\big)\eqno(9)$$
In practice, for  $x\leq n\leq N$,\ $[a]_n$ and $[b]_n$ are both of the 
order of unity, (unless $m$ is nearly 1, in which case $[a]_n,[b]_n
\approx 0$). On the other hand, recalling that $m_R\geq 1$, it can be 
proved that $|{1-m^2\over 1+m^2}|<2$, therefore a good enough estimate 
is
$$[a]_n,\ \ [b]_n\approx 1\eqno(10)$$
Using this and the asymptotic values (8), it can be shown that the 
truncation error in $Q_e$ is bounded by
$$\delta Q_e \leq |a_N|\eqno(11)$$
The proof is presented in Appendix I where it is  shown  
that the series $\sum^\infty_{n=N+1}|a_n|$ converges faster than some 
geometric series. Let us note that what actually appears in $Q_e$ is 
Re$\,a_n$, not $|a_n|$, therefore the bound (11) will usually be 
conservative. This is especially true for small $m_I$ because in this 
case Re$\,a_n\sim|a_n|^2$ (i.e. $Q_e\sim Q_s$) and $|a_n|^2\ll|a_n|$ for 
$n>N$.

Let us now find a criterion for choosing the number $N$ of partial waves 
that should be taken into account. For this purpose let $\epsilon$ be the 
error allowed in the calculation, and let us take $\delta Q_e$ as a 
typical quantity in the problem. Then $N$ should be taken so that 
$$\delta Q_e\leq\epsilon\eqno(12)$$
Taking the quantity $Q_e$ has the advantage of being simple and also 
that $\delta Q_s\leq\delta Q_e$, because $|a_n|^2<{\rm Re}\,a_n$ (i.e. 
$Q_s\leq Q_e$ for each partial wave). Other interesting quantities, such 
as the scattering amplitudes, have similar convergence properties as 
$Q_e$ and $Q_s$.

Putting together the bound (11), the criterion (12) and the estimate 
(10) we find the following prescription
$$\bigg|{\psi_N(x)\over\zeta_N(x)}\bigg| \leq \epsilon\eqno(13)$$
In order to find something more convenient let us make use of the 
Wronskian identity (ref. 4, p.439)
$$\psi_n(x)\zeta_{n-1}(x)-\psi_{n-1}(x)\zeta_n(x)={\rm i},\eqno(14)$$
and the asymptotic values of $A_n(x)$ and $B_n(x)$. In this way we 
obtain (within approximations keeping the order of magnitude)
$$\psi_n(x)\zeta_n(x)\approx-{\rm i}F_n^{-1}(x),\eqno(15)$$
This allows us to remove $\psi_n(x)$  from (13) and finally we obtain 
the prescription for $N$
$$|{\rm Im}\,\zeta_N(x)|\geq \sqrt{1\over\epsilon}\,,\eqno(16)$$
which has been written in a form convenient for being checked while 
$\zeta_n(x)$ is being computed by upward recurrence. In getting (16) we 
have neglected a factor $F_N(x)$ from (15) because by doing so $N$ 
may increase at most by one unit (recall that 
$\zeta_n(x)/\zeta_{n-1}(x)\approx F_n(x)$). Also we have used that 
Re$\,\zeta_n(x)=\psi_n(x)$ is negligible as compared to Im$\,\zeta_n(x)$ in 
the exponential region.

It is remarkable that the value of $N$ obtained from (16) for 
$\epsilon=10^{-8}$ is virtually identical to the standard prescription
$N=x+c\,x^{1/3}+1$, with $c=4.3$. It is shown in Appendix II that it 
must be so using asymptotic expansions for $\zeta_n(x)$, and also how to  
modify $c$ if some other precision $\epsilon$ is desired. To know $N(x)$ 
in advance is necessary if the computer code is to be 
vectorized${}^{10,11}$.

{\bf 3. NUMERICAL ERROR AND UPWARD RECURRENCE}

In this section we shall discuss the propagation of numerical error 
through the calculation.

It is known that the determination of $\psi_n(z)$ by upward recursion is 
intrinsically unstable (see e.g. ref. 5). Let us clarify this 
point.\footnote{*}{We thank one of the referees for providing us with a 
simpler proof of this statement.} For the sake of simplicity let us assume 
that the numerical error is coming from the initial values
$$\tilde\psi_0(z)=\psi_0(z)+\epsilon_0,\ \ \ 
\tilde\psi_1(z)=\psi_1(z)+\epsilon_1\eqno(17)$$
but the recursion itself is free of roundoff error, i.e.
$$\tilde\psi_{n+1}(z)=F_n(z)\tilde\psi_n(z)-\tilde\psi_{n-1}(z)\eqno(18)$$ 
$\epsilon_0,\epsilon_1$ being small numbers depending on the precision 
of the computer, and $\tilde\psi_n(z)$ being the numerical sequence that 
is actually obtained instead of the exact one, $\psi_n(z)$. Subtracting
the exact recursion for $\psi_n(z)$ from (18) we find
$$\delta\psi_{n+1}(z)=F_n(z)\delta\psi_n(z)-\delta\psi_{n-1}(z)\eqno(19)$$ 
where $\delta\psi_n(z)=\tilde\psi_n(z)-\psi_n(z)$ is the error in our 
numerical sequence. Any sequence satisfying the recurrence relation (4) 
is a linear combination of $\psi_n(z)$ and $\zeta_n(z)$, therefore
$$\delta\psi_n(z)=\eta\psi_n(z)+\eta^\prime\zeta_n(z)\eqno(20)$$
The small numbers $\eta,\eta^\prime$  are directly related to 
$\epsilon_0,\epsilon_1$ through eqn. (17), namely
$$\eqalign{\eta &= {\rm i}(\epsilon_0\zeta_1-\epsilon_1\zeta_0) \cr
\eta^\prime &= -{\rm i}(\epsilon_0\psi_1-\epsilon_1\psi_0)\cr}
\eqno(21)$$
Recalling now that $\zeta_n(z)$ diverges for large $n$ we conclude that 
the absolute error in $\tilde\psi_n(z)$ will eventually blow up. More 
generally, if the recursion itself is not exact due to computer roundoff 
error, $\tilde\psi_n(z)$ is rather given by
$$\delta\psi_n(z)=\eta_n\psi_n(z)+\eta_n^\prime\zeta_n(z)\eqno(22)$$
where $\eta_n,\eta^\prime_n$ are of the order of the roundoff error or
the initial values error, whichever the largest. In any case the 
conclusion is still that $\delta\psi_n(z)$ is small for small $n$ (or 
while $n$ is in the oscillating regime for $z$ nearly real), but blows 
up when $n$ enters in the exponentially increasing regime of 
$\zeta_n(z)$. Since $\psi_n(z)$ itself goes to zero in the 
exponential regime, 
$\tilde\psi_n(z)$ has less and less correct figures at each step.

We can extract some corollaries from the previous discussion:

1) The upward recursion is always unstable for computing $\psi_n(z)$ for 
large $n$, depending on $z$. The error $\delta\psi_n(z)$ grows as 
$|\zeta_n(z)|$. On the other hand the upward recursion is perfectly 
stable for computing $\zeta_n(x)$ for any value of $n$. This is because 
$\delta\zeta_n(x)$ still grows as $|\zeta_n(x)|$, therefore the relative 
error in $\zeta_n(x)$ is kept small. Note however that the 
relative error in the quantity Re$\,\zeta_n(x)=\psi_n(x)$ is not at all 
small.

2) A downward recursion is stable for computing $\psi_n(z)$, because 
$|\zeta_n(z)|$ is either slowly changing (in the oscillating regime) or 
quickly decreasing with decreasing $n$ (in the exponential regime). This
allows for taking even very rough estimates for the initial 
values of $\psi_n(z)$ in the downward recursion and the ratio 
$\tilde\psi_{n-1}(z)/\tilde\psi_n(z)$ will still quickly approach the 
exact value $A_n(z)$. On the other hand, a downward recursion is not 
appropriate 
for computing $\zeta_n(x)$ or the ratio $B_n(x)$ if it starts in the 
exponential regime.

Now let us study the influence of the numerical error on the $a_n,b_n$ 
coefficients, and hence on $Q_e$ if an upward recursion is used to 
compute $\psi_n(x)$. In this analysis $\zeta_n(x)$ and $B_n(x)$ are 
assumed to be exact due to previous considerations. On the other hand 
$A_n(y)$ is also assumed to be exact. The effect of using approximate 
values of $\psi_n(y)$ will be considered later. We can make the 
discussion for $a_n$. Similar conclusions will hold for $b_n$. Eqn. 
(6) can be rewritten as 
$$a_n=  {\psi_n(x)\over\zeta_n(x)} f(A_n(x)),\eqno(23)$$
where only the $A_n(x)$ dependence is shown explicitly as it is the 
only relevant one for error analysis. The relative error in $a_n$ will 
be given by
$${\delta a_n\over a_n}\approx {\delta \psi_n\over\psi_n} + 
{f^\prime\over f}{\delta A_n\over A_n}. \eqno(24)$$
Recalling the definition (7), the relative error in $A_n$ can be 
estimated to be of the same order of magnitude as that of $\psi_n$, and 
taking into account that $f$ is a smooth function of the order of unity
(cf. eqn. (10)), one gets the estimate
$$\delta a_n\approx a_n{\delta\psi_n\over\psi_n}\approx 
a_n\eta^\prime{\zeta_n\over\psi_n} = \eta^\prime f\approx \eta^\prime\ . 
\eqno(25)$$
where use has been made of eqn. (22) and $\eta^\prime$ is some typical 
value of $\eta^\prime_n$.

This means that the absolute error in $a_n$ or $b_n$, remains roughly 
constant throughout the computation. 
Of course eqn. (24) holds only for small 
$\delta\psi_n$, but this is guaranteed as $N$ is of the order of $x$ and 
so the recurrence does not go deep inside the exponential region. The 
important consequence of eqn. (25) is that the upward recursion can be 
used to obtain $\psi_n(x)$ because the error introduced is of the order 
of the roundoff error (see however the comment at the end of Section 6).
Let us note that this fact is consistent with 
available algorithms for doing Mie calculations, where $\psi_n(x)$ and 
$\zeta_n(x)$ are always computed by upward recursion (e.g. refs. 5,11).

Let us consider now the effect of the numerical error coming form 
$\psi_n(y)$. We have argued before that an upward recursion would not be 
appropriate for computing $\psi_n(z)$ in general, however we have just 
shown that it can be used in the case of $\psi_n(x)$. The reason for 
this was that the relative error in $\psi_n(x)$ grew as 
$\zeta_n(x)/\psi_n(x)$ but the quantities $a_n$ and $b_n$ themselves
converged to zero as $\psi_n(x)/\zeta_n(x)$. Both factors cancel 
rendering $\delta a_n$ and $\delta b_n$ bounded. We cannot apply a 
similar argument to $\delta\psi_n(y)$ and therefore an upward recursion 
is not reliable to compute $\psi_n(y)$ for arbitrary $y$. We can 
consider two limiting cases
\item{a)} $m_I=0$. In this case $y$ is real and greater than $x$, thus the 
instability in $\psi_n(y)$ starts only after that in $\psi_n(x)$, therefore
the upward recursion can be used. 
\item{b)} Large $m_I$. From the initial values${}^4$
$$\eqalign{
\psi_0(z)  &= \sin(z)\ ,\phantom{{\rm i}\exp(-{\rm i}z)}\ 
\psi_1(z)={1\over z}\sin(z)-\cos(z)\cr
\zeta_0(z) &= {\rm i}\exp(-{\rm i}z)\ ,\phantom{\sin(z)}\ 
\zeta_1(z) =
\bigg({{\rm i}\over z}-1\bigg)\exp(-{\rm i}z) \cr}\eqno(26)$$
\item{}one can see that $\psi_n\sim\exp(m_Ix),\ \zeta_n\sim\exp(-m_Ix)$, for 
small $n$, thus $\psi_n$ is much larger than $\zeta_n$. On the other 
hand $\epsilon_{0,1}$ are related to the computer precision, typically 
$\epsilon_{0,1}\sim r\psi_{0,1}$ with $r\approx 10^{-16}$ in double 
precision. Upon substitution in (21) we find that $\eta$ is small but
$\eta^\prime\sim r\exp(2m_Ix)$ which is not necessarily small. The 
relative error in  $\tilde\psi_n(z)$ goes as 
$${\delta\psi_n(z)\over\psi_n(z)}\approx 
r\bigg|{\psi_0(z)\over\zeta_0(z)}\bigg|\,
\bigg|{\zeta_n(z)\over\psi_n(z)}\bigg|\eqno(27)$$
\item{}For small $n$ the relative error is small, of the order of $r$, however 
for $n\sim|z|$, where $\psi_n$ and $\zeta_n$ are of the order of unity, 
the relative error is $r|\psi_0/\zeta_0|\sim r\exp(2m_Ix)$ which is 
large for large $m_I$. Therefore the upward recursion is not stable in 
this case.

To summarize, the upward recursion to compute $\psi_n(y)$ can be used if 
$m_I$ is small enough but becomes unstable for large $m_I$.
We have not analyzed in any detail in which cases the upward recursion 
for $\psi_n(y)$ is reliable, therefore we shall only consider downward 
recurrences for this quantity. See however refs. 10,11 for an extensive 
analysis of this problem through computer experiments.
Noting that all we need is the ratio $A_n(y)$, for $1\leq n\leq N$, we 
can use the downward recursion
$$A_n(y)=F_n(y)-{1\over A_{n+1}(y)}\ . \eqno(28)$$
Computing the initial value $A_N(y)$ requires some algorithm such as 
that of Lentz$^{17}$ or the one we present in the next section. Let us 
estimate now the precision required in $A_N(y)$ in order not to 
introduce an error in $Q_e$ larger than the prescribed precision 
$\epsilon$. By arguments similar to those used for $\psi_n(x)$, we have
$${\delta a_n\over a_n}\approx{\delta A_n(y)\over A_n(y)} \eqno(29)$$
where $\delta a_n$ is the error introduced by $\delta A_n(y)$.
Given that the downward recursion is stable we can assume that
$$\bigg|{\delta a_n\over a_n}\bigg| \leq \bigg|{\delta A_N(y)\over A_N(y)}
\bigg|\ \ \ \ \ \ {\rm for }\ \ \  n\leq N \eqno(30)$$
Using this relationship one gets for the numerical error in $Q_e$ 
$$\delta Q_e\approx {1\over x^2}\sum^N_{n=1}(2n+1)\delta a_n 
\leq Q_e\bigg|{\delta A_N(y)\over A_N(y)}\bigg|. \eqno(31)$$
Therefore the numerical error from $A_n(y)$ will be under control by imposing
$$\bigg|{\delta A_N(y)\over A_N(y)}\bigg|
\leq {\epsilon\over  Q_e}. \eqno(32)$$
Let us note that this criterion will be conservative in general. An 
exception would be the case of $y$ being real and bigger than $N$. In this 
case the recurrence (28) has no healing properties (for it 
already starts in the oscillatory regime) and hence the equal sign is 
reached in (30).

{\bf 4. INITIALIZATION OF THE DOWNWARD RECURRENCE}

In this section we present a new method to compute $A_N(z)$, of 
similar efficiency to that due to Lentz$^{17}$ (actually ours needs one 
multiplication less at each step). This method has the advantage of being able 
to implement a precision condition as that in eqn. (32), hence
controlling the required precision in $A_n(y)$.

Let $X_n(z)$ and $Y_n(z)$ be two 
sequences satisfying the recurrence (4) for some value of $z$
(the dependence on $z$ is irrelevant here). Then they will satisfy
 the Wronskian identity
$$C = X_nY_{n+1}-X_{n+1}Y_n\eqno(33)$$
where $C$ is independent of $n$. We can rewrite it as a difference 
equation
$$C = Y_nY_{n+1}\bigg\{\bigg({X\over Y}\bigg)_n - \bigg({X\over 
Y}\bigg)_{n+1}\bigg\}, \eqno(34)$$
and solve it in $X_n$
$$X_n = DY_n+CY_n\sum^\infty_{k=n}(Y_kY_{k+1})^{-1},\eqno(35)$$
$D$ being a constant. To write (35) we have assumed that $Y_n$ is a 
sequence going to infinity for large $n$, which is true for almost any 
solution of the recurrence (4). If we take $Y_n$ as a fixed sequence and
regard $C,D$ as free parameters, then $X_n$ is the most general 
solution of the recurrence relation (4). In particular for $D=0$,\ 
$X_n$ goes to zero as $n$ goes to infinity, as a consequence it must be
proportional to $\psi_n$,
$$\psi_n(z)=C(z)Y_n(z)\sum^\infty_{k=n}(Y_k(z)Y_{k+1}(z))^{-1}\eqno(36)$$
The constant $C$ cancels after computing the ratio $A_n(z)$
$$A_n(z)=Y_n^{-1}\big\{Y_{n-1} + 
Y^{-1}_n\big[\sum^\infty_{k=n}(Y_kY_{k+1})^{-1}\big]^{-1}\big\}.\eqno(37)$$
Finally, a simpler formula can be obtained for $A_N(z)$ by choosing as 
starting values for the sequence $Y_n$
$$Y_{N-1}=0\ , \ \ \ \ Y_N=1\eqno(38)$$
$$A_N(z)=\big[\sum^\infty_{k=N}(Y_k(z)Y_{k+1}(z))^{-1}\big]^{-1}.\eqno(39)$$
About the convergence of the series in (39), we note that it is very 
fast when $Y_k$ enters in its exponential regime. Note that 
for real $y$ the convergence begins only after $k\geq 
y$. A similar conclusion was reached by other authors$^{11}$ in Lentz's 
method which basically follows the same principle as ours and so has 
similar convergence properties. 

The sequence in eqn. (39)
must be truncated at some value $k=M$ in such a way as to fulfill the 
requirement (32). This can be easily done by noting that the error 
introduced in $A^{-1}_N(y)$ is of the order of the last term taken 
into account (this follows from $|Y_k/Y_{k-1}|\approx |F_k|> 2$ for large 
$k$),
$$\delta A^{-1}_N\approx\big(Y_MY_{M+1}\big)^{-1}.\eqno(40)$$
On the other hand we should require
$$|\delta A^{-1}_N(y)|\approx\bigg|A^{-1}_N{\delta A_N\over 
A_N}\bigg|\leq\bigg|{1\over F_N(y)}{\epsilon\over Q_e}\bigg|\eqno(41)$$
where we have made use of eqns. (8) and (32). Recall now 
that for $x\geq 1$, $F_N$ and $Q_e$ are of the order of unity 
whereas for $x\ll1$ the product of $F_NQ_e$ is still of the order of 
unity, therefore the final criterion to truncate (39) is
$$\big|\big(Y_M(y)Y_{M+1}(y)\big)^{-1}\big|\leq\epsilon.\eqno(42)$$
To finish this section we shall show how to avoid ill-conditioning in 
(39), which will appear if $Y_k$ gets too near to zero for some 
value of $k$. To do this we can use the recurrence relation (4) to write
$${1\over Y_{k-1}Y_k} + {1\over Y_kY_{k+1}} = {Y_{k-1} + Y_{k+1}\over
Y_{k-1}Y_kY_{k+1}} = {F_k\over Y_{k-1}Y_{k+1}}, \eqno(43)$$
which is well behaved even for $Y_k=0$.

{\bf 5. COMPUTATIONAL ALGORITHM}

Using the previous ideas, we have developed a computational 
algorithm which we shall briefly describe now. The input is $x, m$ 
and $\epsilon$ and the main output are the coefficients $a_n$ and $b_n$, 
and $N$. To start with, analytic expressions for $\zeta_0(x)$ and $\zeta_1(x)$ 
are taken to initiate an upward recurrence for $\zeta_n(x)$. This 
quantity is
kept in a (complex) array variable. The recurrence stops when the condition 
(16) 
is fulfilled, providing the value of $N$. The quantities 
$\psi_n(x)$ are automatically obtained as the real part of $\zeta_n(x)$. 
As a second step, $A_N(y)$ is computed using eqns. (38), (39) and (42). 
Here we note that from a computational point of view an equivalent form 
of (42) is more convenient, which consist in doing the check for the 
absolute values of the real and imaginary parts. This is much faster 
than computing the modulus of a complex number.

Then a downward recurrence is performed for $A_n(y)$, eqn. (28), until 
$n=1$. Simultaneously, $a_n$ and $b_n$ are computed using $\zeta_n(x)$ and 
$A_n(y)$. The quantities $Q_s$ and $Q_e$ can then be computed. We have 
not developed any especial algorithm for computing the scattering amplitudes 
$S_1$ and $S_2$. To do this efficiently see for instance ref. 11.

The criteria developed above are intended to be robust, hence they are
rather conservative. As a consequence the error in $Q_e$ 
is smaller than the prescribed precision $\epsilon$. This is 
especially true for small values of $x$, whereas for $x\gg 1$, about two 
more figures than expected are obtained. We point out also that $Q_s$ is 
always obtained as accurately as $Q_e$ or more. This fact was expected 
because the criteria were stated for $|a_n|$ while $Q_s$ goes as 
$|a_n|^2$ which converges faster.

{\bf 6. RESONANT TERMS IN THE MIE SERIES}

Let us recall that after eqn. (7) we stated a smoothness assumption 
for the quantities $[a]_n,[b]_n$, namely that they are nearly constant 
in the $x$ exponential regime and do not play any role in the 
convergence of the Mie series, which is only controlled by the ratio 
$\psi_n(x)/\zeta_n(x)$. In particular this assumption implied that the 
highest partial wave with a relevant contribution is independent of $m$ 
(cf. eqn. (16)). In other words, $N$ is a function of $x$ only. This 
result is also supported numerically, (see for instance refs. 10,11). 
Therefore it was a surprise for us to discover that strictly speaking 
such a statement must be false. Moreover, for any choice of $N$ as a 
function of $x$ only, and for any prescribed value of n,\  $n>N$, one can 
always pick a value of $m$ (in fact infinitely many of them) in such a 
way that the $n$-th term in the Mie series is not negligible, for 
instance one can make $a_n=1$. The consequence of this that in order to 
guarantee that the numerical value of $Q_e$ is correct within some 
prescribed precision, $N$ should depend on $m$ as well as on $x$.

In order to clarify the point let us consider the worst case, which is 
also the simplest, namely $m_I=0$, i.e. $y$ real. This is the only case 
in which $|a_n|$ or $|b_n|$ can reach the value 1. The point can be made 
for $a_n$: recalling that for $z$ real Re$\,\zeta_n(z)=\psi_n(z)$, eqn. 
(2) can be rewritten as
$$\eqalign{a_n &= {{\rm Re}\,D_n\over D_n} \cr
 D_n &= 
x\zeta_n(x)\psi^\prime_n(y)-y\zeta^\prime_n(x)\psi_n(y)\cr}\eqno(44)$$
where $D_n$ is a complex quantity. Obviously $a_n=1$ if and only if 
$${\rm Im}\,D_n = 0\ .\eqno(45)$$
Let us regard $x$ and $n$ as given and look for solutions of (45) in 
the variable $y$. The equation can be rewritten as
$${1\over y}{\psi_n^\prime(y)\over\psi_n(y)} =
{1\over x}{{\rm Im}\,\zeta_n^\prime(x)\over{\rm Im}\,\zeta_n(x)} 
\eqno(46)$$
In the interval $y>n$, $\psi_n(y)$ is a real oscillating function of $y$ 
with infinitely many zeroes. Between two zeroes of $\psi_n(y)$, the 
l.h.s. of eqn. (46) takes every real value, therefore there are 
infinitely many solutions to our equation for any values of $x$ and $n$, 
no matter how large is $n$ as compared to $x$. For these values of 
$x,m$, and $n$,\ \  $a_n$ will not at all be negligible.

Let us now show that these resonances do not occur for unrealistic 
values of $m$. Typically (and asymptotically for large $y$) the distance 
between two consecutive zeroes of $\psi_n(y)$ is of the order of $\pi$, 
therefore for given $x$ and $n$  the lowest resonant value of $m$ will 
occur near the interval $({n\over x},{n+\pi\over x})$ approximately. For 
large $x$ this happens for $m$ near to unity, and all the other resonant 
values will follow at a distance of about $\pi/x$ from each other.

From a rigorous point of view these findings would invalidate the 
estimates (10) and their consequences. They would also invalidate any 
algorithm in which $N$ depends on $x$ only, namely every existent 
algorithm known to us. In fact the only practical way to make sure that 
the resonant partial waves have been accounted for would be to take $N$ 
greater than $y$ in order to guarantee that $\psi_n(y)$ has no zeroes 
for $n>N$.

Nevertheless it is clear that in practice the existent algorithms to do Mie 
scattering calculations work. To account for this fact we 
should consider not only the existence of resonant partial waves but also 
their width. Let us show that for sensible choices of $N$ (as a function 
of $x$) and for $n>N$ the resonances are so narrow that  they will not 
normally show up. Let $y_0$ be one the values of $y$ such that $a_n=1$. A 
look to eqn. (44) shows that for generic $y$, Re$\,D_n$ goes as 
$\psi_n(x)$ whereas $D_n$ goes as $\zeta_n(x)$, therefore $a_n$ is very 
small. However for the especial value $y_0$ there is a cancellation 
between two huge numbers in Im$\,D_n$, leaving $a_n$ of the order of 
unity. The range of values of $y$ for which a partial cancellation takes 
place is related to the slope of $D_n$ in $y=y_0$, namely
$$\Gamma \approx \bigg|{D_n\over D^\prime_n}\bigg|_{y=y_0} = 
 \bigg|{{\rm Re\,}D_n\over D^\prime_n}\bigg|_{y=y_0} \approx 
\bigg|{\psi_n(x)\over\zeta_n(x)}\bigg|\ . \eqno(47)$$
Where $D^\prime_n = {\rm d}D_n/{\rm d}y$. In other words, if $N$ is 
large enough only by a very careful choice of $m$ or $x$ can one find 
one these resonant contributions. More precisely, recalling eqn. (13), 
we can see that $m$ or $x$ should be fine tuned at least with a precision 
$\epsilon$ in order to pick a resonant term for some $n>N$. On the other 
hand, except for these rare cases, $a_n,b_n$ are indeed small and of the 
order of $\psi_n(x)/\zeta_n(x)$, therefore our analysis applies.
If $m$ is allowed to be complex, a more involved analysis would be 
needed, but we expect that the conclusion would not differ.

Let us finally note another consequence of the resonant terms on the 
calculation, even when they are taken into account. For one of these terms 
the quantity $f$ in eqn. (23) is no longer of the order of unity, on the 
contrary it is rather large, and the last step in eqn. (25) cannot be taken.
This means that a resonant term amplifies the error due to $\psi_n(x)$. 
The cure is simply to compute $\psi_n(x)$ by downward recursion for 
$x < n < N$. This has in fact been observed in selected 
quantities such as the backscattering efficiency for suitable values of 
$x$ and $m$ (Ref. 5).

{\bf 7. CONCLUSIONS}

In this paper we have addressed several points relevant to Mie 
scattering calculations. To be specific:

a) We have estimated the error introduced in the calculation by 
truncating the Mie series, thereby finding a prescription for choosing 
$N$. We have found that in the generic case $N$ depends on $x$ only.

b) The possible instabilities in the recursions used to compute $\psi_n$ 
and $\zeta_n$ have been analyzed. We have found that upward recursion
is always unstable for computing $\psi_n(z)$ if $n$ is large enough. 
However it can be used to compute $\psi_n(x)$ in Mie calculations. As a 
matter of fact $\psi_n(x)$ is computed in this way in nowadays 
available algorithms. We have also found that upward recursion can be 
used for $\psi_n(y)$ if $m_I$ is small enough, but no criterion is 
given for how small $m_I$ should be.

c) A criterion has been established for the allowed error in 
$\psi_{n-1}(y)/\psi_n(y)$.

d) A new method to compute $\psi_{n-1}(y)/\psi_n(y)$ is presented which 
is efficient and allows for controlling the error and removing ill-conditioning.

e) It has been shown the existence of resonant terms in the Mie series 
which can also appear for $n>N$. Strictly speaking the existence of 
these terms invalidates any algorithm in which $N$ is a function of $x$ 
only. However we have also shown that those resonant terms are extremely 
rare, namely they appear with a probability of the order of $\epsilon$.

A specific algorithm is also described. It is meant to be robust and 
efficient for a wide range of size parameters and refractive indices.
 With 
this algorithm we have written the computer program LVEC-MIE$^{18}$,
which is available both in single and double precision
contacting V.E. Cachorro.

\eject
{\bf APPENDIX I}

Let us justify the bound (11). To do so we shall study the convergence 
rate of the terms left out in the series, $n>N$. In this region we can 
make use of the estimate (10),
$$\eqalign{
\delta Q_e &= {2\over x^2}\sum^\infty_{n=N+1} (2n+1) {\rm Re\,}(a_n + 
b_n)\cr
&\leq {2\over x^2}\sum^\infty_{n=N+1} (2n+1) \big(|a_n| + |b_n|\big)\cr
&\approx {8\over x^2}\sum^\infty_{n=N+1} n 
\bigg|{\psi_n(x)\over\zeta_n(x)}\bigg|\cr
&= {8\over x^2}\sum^\infty_{n=N+1} n 
\bigg|{B_{n}(x)\over A_{n}(x)}\,{B_{n-1}(x)\over A_{n-1}(x)}\dots
{B_{N+1}(x)\over 
A_{N+1}(x)}\bigg|\,\bigg|{\psi_N(x)\over\zeta_N(x)}\bigg| \ 
.\cr}\eqno({\rm I.1})$$
Now making use of (8) and recalling that $F_n(x)$ is a monotonically
increasing function of $n$, we obtain
$$\eqalign{\delta Q_e &\leq {8\over x^2}\sum^\infty_{n=N+1} n 
{1\over F^2_{n}(x)}\,{1\over F^2_{n-1}(x)}\dots{1\over F^2_{N+1}(x)}
\,\bigg|{\psi_N(x)\over\zeta_N(x)}\bigg|\cr
&\leq {8\over x^2}\sum^\infty_{n=N+1} n \big(F_N(x)\big)^{2(N-n)}
\,\bigg|{\psi_N(x)\over\zeta_N(x)}\bigg|\cr
&= {8\over x^2}\bigg({N\over F^2_N(x)-1} + {F^2_N(x)\over 
\big(F_N^2(x)-1\big)^2}\bigg)\,|a_n| \cr}\eqno({\rm I.2})$$

For small $x$, $N=1$ and $F_N(x)$ is large, hence
$$\delta Q_e \leq 2 |a_n|\eqno({\rm I.3})$$
on the other hand, for large $x$, $N\approx x$ and $F_N(x)\sim 2$,
$$\delta Q_e \leq {8\over 3}{1\over x} |a_n|\ .\eqno({\rm I.4})$$
In both cases eqn. (11) is valid (up to factors of the order of unity).

\eject
{\bf APPENDIX II}

In order to know in advance the value of $N$ that will be obtained from 
the prescription (16) for given $x$ and $\epsilon$, let us recall that
Im$\,\zeta_n(x)=\sqrt{\pi x/2}\,Y_{N+{1\over 2}}(x)$,\ $Y_\nu(z)$ 
being the Bessel function of the second kind. Let $\nu$ and $c$ be 
defined by
$$\eqalign{
N &= \nu-{1\over 2} \cr
x &= \nu-c\,\nu^{1/3} \ . \cr}\eqno({\rm II.1})$$
Note that for large $\nu$, eqn. (II.1) can be inverted to give $N\approx 
x+c\,x^{1/3}$. Now we can make use of the leading order term in the 
asymptotic expansion of $Y_\nu$ for large $\nu$ and fixed $c$, 
ref. 4, p. 367:
$${\rm Im\,}\zeta_N(x)\sim -\sqrt{\pi}\bigg({\nu\over 
2}\bigg)^{1/6}\,{\rm Bi}\big(2^{1/3}c\big)\ ,\eqno({\rm II.2})$$
where Bi$(z)$ is the Airy function of the second kind, ref. 4, p. 446. 
This function is given by
$${\rm Bi}(z) = z^{-1/4}\,f(z)\exp({2\over 3}z^{3/2})\ ,\eqno({\rm II.3})$$
where $f(z)$ is nearly constant for $z>1$ with 
$f(z)\approx 1/\sqrt{\pi}$,\  ref. 4, p. 449. Thus
$$\bigg|{\rm Im\,}\zeta_N(x)\bigg|\approx \bigg({\nu\over 
2\sqrt{2}}\bigg)^{1/6}\,c^{-1/4}\exp\big({1\over 3}(2c)^{3/2}\big)\ . 
\eqno({\rm II.4})$$
The right hand side of (II.4) has a very strong dependence on $c$ 
whereas it depends very smoothly on $\nu$. Actually 
$(\nu/2\sqrt{2})^{1/6}$ is of the order of unity for  $\nu=1$ up to 
$10^5$. Therefore using eqn. (16), $c$ will be determined by 
$\epsilon$. We find that $c=4.3$ corresponds to $\epsilon=10^{-8}$. 
Other values are $c=4.0,\ \epsilon=10^{-7}$, and $c=5.0,\ 
\epsilon=10^{-10}$, computed for $\nu=100$ in (II.4). 
\vfill
\eject
\null 
\medskip

{\bf REFERENCES}

\item{1.} H. C. van de Hulst, {\it Light Scattering by Small 
Particles}, John Wiley, N. Y. 1957.
\item{2.} M. Kerker, {\it The Scattering of  Light and Other 
Electromagnetic Radiation}, Academic Press. N. Y., 1969.
\item{3.} C. F. Bohren and D. R. Huffman, {\it Absorption and 
Scattering of Light by Small Particles}, Wiley Interscience, N. Y. 1983.
\item{4.} M. Abramowitz and I. A. Stegun ed., {\it Handbook of Mathematical 
Functions with Formulas, Graphs and Mathematical Tables}, Dover Pub. 
Inc., N. Y., 1965.
\item{5.} J. V. Dave, {\it Subroutines for Computing the Parameters of 
Electromagnetic Radiation Scattered by a Sphere}, Report No. 320-3237, 
IBM Scientific Center, Palo Alto, California, USA, 1968.
\item{6.} J. V. Dave, {\it Scattering of Electromagnetic Radiation by 
Large Absorbing Spheres}, IBM J. Res. Develop., Vol.13, 1302-1313, 1969. 
\item{7.} W. Blattner, {\it Utilization Instruction for Operation of the 
Mie Programs on the CDC-6600 Computer at AFCRL}, Radiation Center 
Associates, Ft. Worth, Texas, Res. Note RRA-N7240, 1972.
\item{8.} G. Grehan and G. Gouesbet, {\it The Computer Program SUPERMIDI 
for Mie Theory Calculation, without Practical Size nor Refractive Index 
Limitations}, Internal Report TTI/GG/79/03/20, Laboratoire de G\'enie 
Chemique Analytique, U. de Rouen, 76130 Mt-St-Aignan (France), 1979. 
Also Private communication.
\item{9.} G. Grehan and G. Gouesbet, {\it Mie theory calculations: new 
progress, with emphasis on particle sizing}, Appl. Opt., Vol. 18, 
3489-3493, 1979.
\item{10.} W. J. Wiscombe, {\it Mie scattering calculations: Advances in 
technique and fast vector speed computer codes}. NCAR Technical Note 
NCAR/TN-140+STR (National Center for Atmospheric Research) Boulder, 
Colorado, 80307, 1979, and private communication.
\item{11.} W. J. Wiscombe, {\it Improved Mie Scattering Algorithms}, 
Appl. Opt., Vol. 19, 1505-1509, 1980.
\item{12.} G. H. Goedecke, A. Miller and R. C. Shirkey, {\it Simple 
Scattering Code Agausx}, in Atmospheric Aerosols: Their Formation, 
Optical Properties and Effects. Ed. A. Deepak, Spectrum Press, Hampton, 
Virginia, 1982.
\item{13.} A. Miller, {\it Comments on Mie Calculations}, Am. J. Phys., 
Vol. 54, 297-297, 1986. Also private communication.
\item{14.} G. W. Kattawar and G. N. Plass, {\it Electromagnetic 
Scattering from Absorbing}
\item{} {\it Spheres}, Appl. opt., vol. 6, 1377, 1967.
\item{15.} D. Deirmendjian, {\it Electromagnetic Scattering on Spherical 
Polydispersion}, Elsevier, N. Y. 1969.
\item{16.} H. Quenzel and H. M\"uller, {\it Optical properties of single 
particles diagrams of intensity, extinction scattering and absorption 
efficiencies}, Wissenschaftliche Mitteilung, n. 34. Metereologisches 
Institut, Universit\"at M\"unchen, 1978.
\item{17.} W. J. Lentz, {\it Generating Bessel Functions in Mie 
Scattering Calculations using Continued Fractions}, Appl. Opt., vol. 15, 
668-671, 1976.
\item{18.} V. E. Cachorro, L. L. Salcedo and J. L. Casanova, {\it 
Programa LVEC-MIE para el c\'alculo de las magnitudes de la teor\'{\i}a 
de esparcimiento de Mie}, Anales de F\'{\i}sica, vol. 85, Serie B, 
198-211, 1989.
\vfill
\eject
\bye